\documentclass[twocolumn,showpacs,preprintnumbers,amsmath,amssymb,groupedaddress,superscriptaddress]{revtex4}
\usepackage[dvips]{graphicx}
\usepackage[dvips]{color}

\begin{document}

\title{Comparison of ${}^{87}$Rb \emph{N}-resonances for $D_1$ and $D_2$
   transitions}

\author{Irina Novikova, David F.\ Phillips, Alexander S.\ Zibrov$^*$,
   Ronald L.\ Walsworth}
\affiliation{Harvard-Smithsonian Center for Astrophysics and
Department of Physics, Harvard University, Cambridge, MA, 02138, USA}
\author{Alexksei V.\ Taichenachev, Valeriy I.\ Yudin}
\affiliation{Institute of Laser Physics SB RAS and Novosibirsk State
University, Novosibirsk, 630090, Russia}
\begin{abstract}
   We report an experimental comparison of three-photon-absorption
resonances (known as
   ``\emph{N}-resonances'') for the $D_1$ and $D_2$ optical transitions
   of thermal ${}^{87}$Rb vapor.  We find that the $D_2$
\emph{N}-resonance has better contrast, a broader linewidth, and a
more symmetric
lineshape than the $D_1$ \emph{N}-resonance. Taken together, these
factors imply
   superior performance for frequency
   standards operating on alkali $D_2$ \emph{N}-resonances, in contrast to
   coherent population trapping (CPT) resonances for which the $D_2$
   transition provides poorer frequency standard performance than the $D_1$
   transition.
\end{abstract}

\date{\today}

\maketitle 

Recently, we demonstrated that three-photon-absorption
resonances known as
``\emph{N}-resonances''\cite{Zibrov-Observation,zibrov05pra,Nlightshifts}
are a promising
alternative to coherent population trapping (CPT)
resonances\cite{vanier05apb,knappe05oe,schwindt04,matsko05,Finland}
for small atomic
frequency standards using thermal alkali vapor. In this
letter, we report an experimental comparison of \emph{N}-resonances for
the ${}^{87}$Rb $D_1$ ($5{}^2\mathrm{S}_{1/2}\rightarrow
5{}^2\mathrm{P}_{1/2}$, $\lambda=795$ nm) and $D_2$
($5{}^2\mathrm{S}_{1/2}\rightarrow 5{}^2\mathrm{P}_{3/2}$,
$\lambda=780$ nm) optical transitions. We find similar
\emph{N}-resonance quality factors for the $D_1$ and $D_2$
transitions, but a significantly more symmetric lineshape for the
$D_2$ transition, which together implies superior performance for a
frequency
   standard using the $D_2$ \emph{N}-resonance. Previous work has
shown that the quality factor of alkali CPT resonances is about an
order of magnitude worse for $D_2$ operation than for
$D_1$.\cite{stahler02,lutwak03} Thus, as current miniature frequency
standards rely on vertical-cavity surface emitting lasers (VCSELs)
that are readily available for the $D_2$ transitions of Rb and Cs but
more
difficult to acquire for the $D_1$ transitions, the results reported
here provide an additional
practical advantage for the \emph{N}-resonance.

\begin{figure}[t]
\centering
\includegraphics[width=0.7\columnwidth]{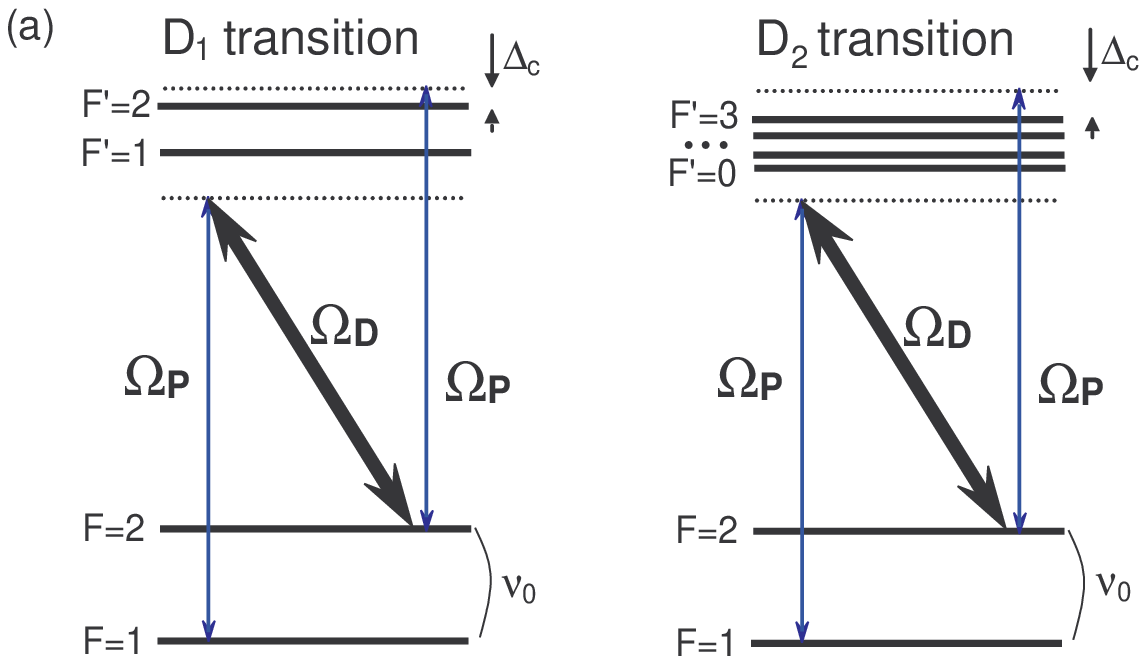}
\includegraphics[width=0.7\columnwidth]{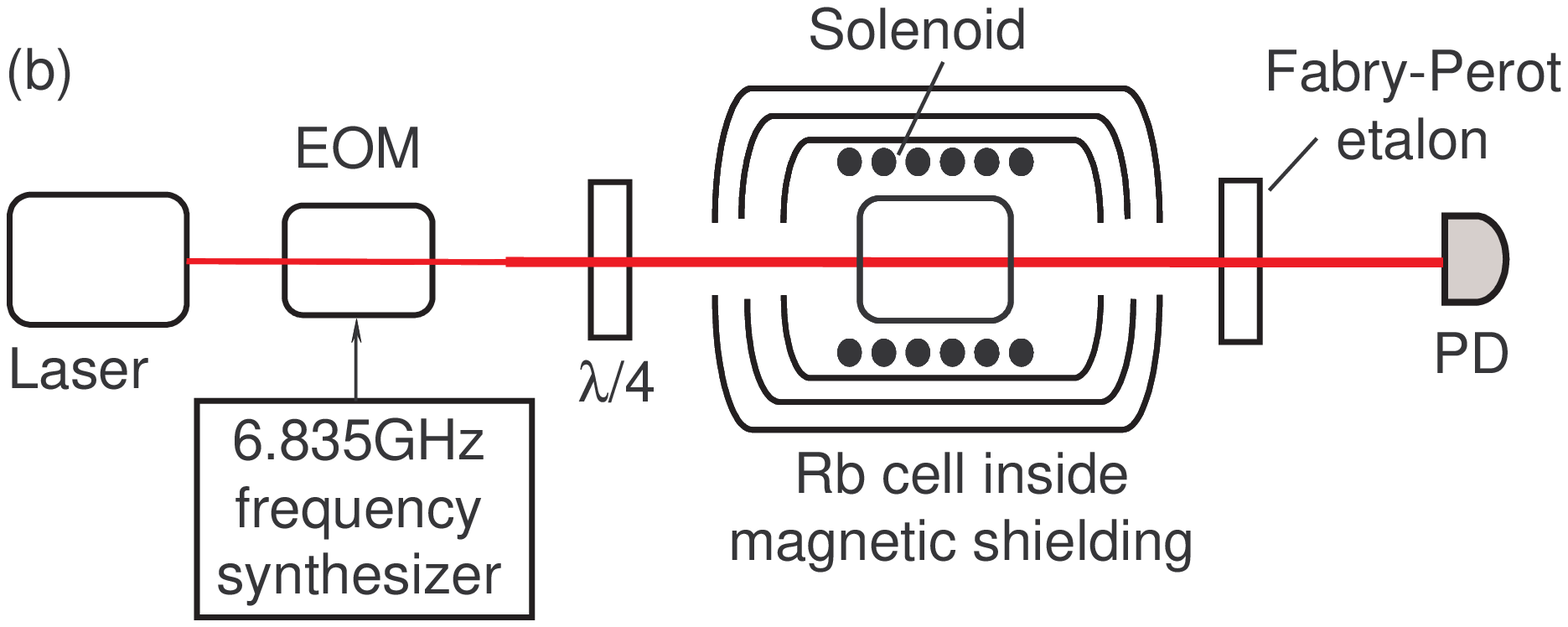}%
\caption{ (a) \emph{N}-resonance interaction scheme for $D_1$ and
$D_2$ transitions of ${}^{87}$Rb. $\Omega_P$ and
   $\Omega_D$ are probe and drive optical fields that create and
   interrogate the \emph{N}-resonance; $\nu_0$ is the
   splitting of the ground-state hyperfine levels $F=1$ and $F=2$; and
   $\Delta_c$ is the one-photon detuning of the probe field from
   resonance between the $F=2$ ground state and the excited state.
(b) Schematic of the experimental setup. See text for abbreviations.}
\label{f.schem}
\end{figure}

An \emph{N}-resonance is a three-photon, two-optical-field absorptive resonance
(Fig.~\ref{f.schem}a). A probe field, $\Omega_P$, resonant with the
transition between the higher-energy hyperfine level of the ground
electronic state  and an electronically excited state, optically pumps
the atoms into the lower hyperfine level, leading to
increased transmission of the probe field through the medium. A drive
field, $\Omega_D$ is detuned from the probe field by the atomic
hyperfine frequency, $\nu_0$. Together, $\Omega_P$ and
$\Omega_D$ create a two-photon
Raman resonance that drives atoms coherently from the lower to the upper
hyperfine level, thereby inducing increased absorption of the probe
field $\Omega_P$ in a narrow resonance with linewidth, $\Delta\nu$,
set by ground-state hyperfine decoherence. Attractive
features of \emph{N}-resonances for atomic frequency standards
include high resonance contrast,
leading-order light-shift cancellation, and less sensitivity than CPT
resonances to high buffer gas
pressures.\cite{zibrov05pra,Nlightshifts}

Better performance for \emph{N}-resonances than CPT resonances on the
$D_2$ transition is expected due to the difference between the resonance
mechanisms.
The CPT transmission maximum appears as a result of optical pumping of
atoms into a non-interacting coherent superposition of two
ground-state hyperfine levels (a ``dark state''). However, a pure
dark state exists only for the $D_1$
transition, thus the amplitude and contrast of CPT resonances are much
higher for $D_1$ operation than for $D_2$.\cite{stahler02}
For \emph{N}-resonances, the three-photon absorptive process does not
require a dark state, with no resultant advantage of $D_1$ over $D_2$.
Furthermore, for circularly polarized light (commonly used in
optically-pumped atomic clocks) resonant with the $D_1$ transition,
some atoms become trapped in the Zeeman state with maximum angular
momentum --- an ``end'' state ( $F = 2, m_F = \pm 2$ for ${}^{87}$Rb) ---
which limits resonance amplitude and contrast on the ground-state
$\Delta$$m_F=0$ hyperfine clock transition.  However, for
$D_2$ operation, the end state is coupled to the excited state through the
cycling transition $F = 2 \rightarrow F^\prime = 3$. In the presence
of strong collisional mixing in the excited state (induced by
alkali-buffer gas collisions), the cycling
transition suppresses optical pumping of atoms into the end state.
Thus we expect higher resonance contrast for \emph{N}-resonances operating on
the $D_2$ transition.

Figure~\ref{f.schem}b shows a schematic of our \emph{N}-resonance
apparatus.  We operated the system under conditions identified in our
previous work to give good frequency standard
performance.\cite{zibrov05pra,Nlightshifts}
We derived the two optical fields $\Omega_P$ and
   $\Omega_D$ by phase modulating the output of an external
cavity diode laser tuned to
either the $D_1$ or $D_2$ transition of ${}^{87}$Rb.  Laser
phase-modulation was performed at the ${}^{87}$Rb ground-state
hyperfine frequency ($\nu_0\simeq 6.8$~GHz) by an electro-optic
modulator (EOM)
driven by an amplified microwave synthesizer phase-locked to a
hydrogen maser.  We used
the $+1$ sideband as the probe field and the zeroth order carrier as
the drive field. We set the laser detuning $\Delta_c$ and EOM
modulation index to
match the conditions for leading-order light-shift
cancellation:~\cite{Nlightshifts} for $D_1$, $\Delta_c\approx+700$
MHz from the $F=2 \rightarrow
F^\prime = 2$ transition; for $D_2$,
$\Delta_c\approx+500$ MHz from the  $F=2 \rightarrow
F^\prime = 3$ transition; in both cases the modulation index $\approx
0.38$, corresponding to a probe/drive intensity ratio of about 19\%.
The laser beam was circularly polarized by a quarter wave
plate ($\lambda$/4) and weakly focused to a diameter of 0.8 mm before
entering a Pyrex cell of length 7 cm and diameter 2.5 cm containing
isotopically enriched ${}^{87}$Rb and 100 Torr of Ne buffer gas
(which induced excited-state collisional broadening $\approx 2$ GHz).
We heated the cell to $\approx 65\,{}^\circ$C, yielding ${}^{87}$Rb
density $\approx4\cdot 10^{11} \text{cm}^{-3}$. We isolated the Rb
vapor cell from external magnetic fields using three layers of
high-permeability shielding, and applied a small ($\approx
10$~mG) longitudinal magnetic field to lift the degeneracy
of the Zeeman sublevels and separate the $F=1$, $m_F=0$ to $F=2$,
$m_F=0$ clock transition (with first order magnetic field
independence) from the $m_F=\pm1$ transitions with first order
magnetic field sensitivity. The transmitted probe field power was
detected by a photodetector (PD); the strong drive field and the
off-resonant lower sideband were filtered from the transmitted laser
beam by a quartz narrow-band
Fabry-Perot etalon (FSR = 20 GHz, finesse = 30) tuned to the frequency of
the probe field.

\begin{figure}
\centering
\includegraphics[width=0.7\columnwidth]{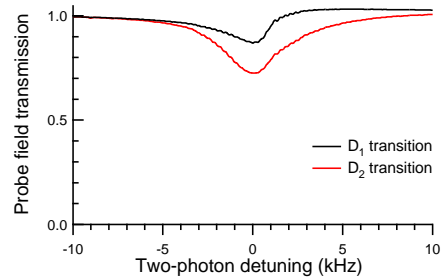}
\caption{
   Example ${}^{87}$Rb \emph{N}-resonances observed for $D_1$ and $D_2$ optical
   transitions. Laser power is 260 $\mu$W, corresponding to an
   intensity of $50$ mW/cm${}^2$. Probe transmission is normalized to
   unity away from two-photon resonance.  }
\label{f.sampleD1D2}
\end{figure}

\begin{figure}[tb]
\includegraphics[width=0.9\columnwidth]{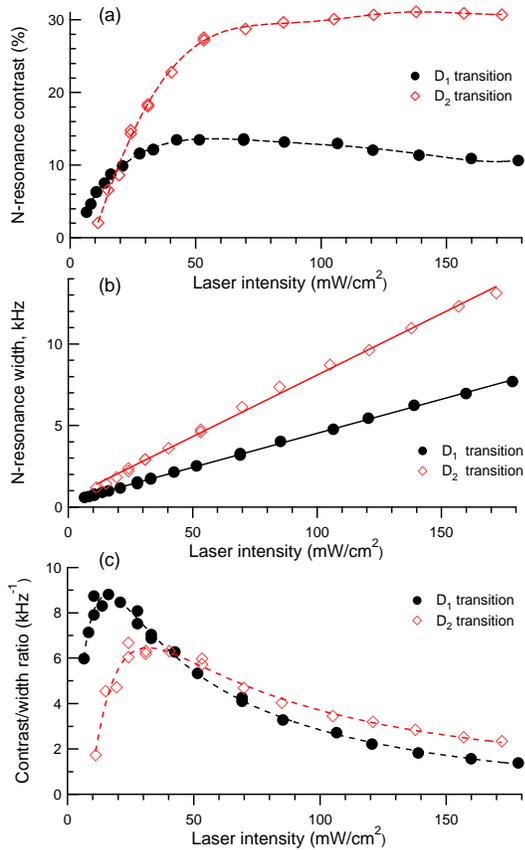}
\caption{
   \emph{N}-resonance (a) contrast $C$, (b) linewidth $\Delta\nu$, and (c)
   quality factor $q=C/\Delta\nu$, as a function of total input laser
intensity, measured on the ${}^{87}$Rb
   $D_1$ (solid circles) and $D_2$ (open diamonds) transitions. Dashed
   lines are to guide the eye.}
\label{f.contrastWidthD1D2}
\end{figure}

Figure~\ref{f.sampleD1D2} shows examples of measured $D_1$ and $D_2$
\emph{N}-resonances under
identical conditions, with the probe
field transmission normalized to unity away from two-photon
resonance. Figure~\ref{f.contrastWidthD1D2} shows the measured
dependence on laser intensity of the \emph{N}-resonance contrast,
linewidth, and quality factor. We define resonance contrast as
$C=1-T_{min}/T_{o}$, where $T_{min}$ and $T_{o}$ are the transmitted
probe field intensities on two-photon resonance and away from
resonance, respectively. The linewidth, $\Delta\nu$, is the measured
full-width-half-maximum (FWHM); and the resonance quality factor is
$q=C/\Delta\nu$. [The shot-noise limit to atomic clock frequency
stability is inversely proportional to the
quality factor; i.e., larger $q$ corresponds to better frequency
stability.~\cite{vanier05apb}] As seen in Figures~\ref{f.sampleD1D2}
and \ref{f.contrastWidthD1D2}, the typical measured
\emph{N}-resonance contrast is significantly larger for $D_2$
operation than for the $D_1$ transition, whereas the linewidth is
broader for $D_2$ than for $D_1$. The differences in contrast and
linewidth largely offset each other, such that the \emph{N}-resonance
quality factor is roughly comparable for the $D_1$ and $D_2$
transitions over a wide range of operating conditions. Note that in
alkali vapor CPT resonances, the quality factor for the $D_2$
transition has been measured to be about an order of magnitude
smaller than for the $D_1$ transition.\cite{stahler02,lutwak03}

\begin{figure}
\centering
\includegraphics[width=0.8\columnwidth]{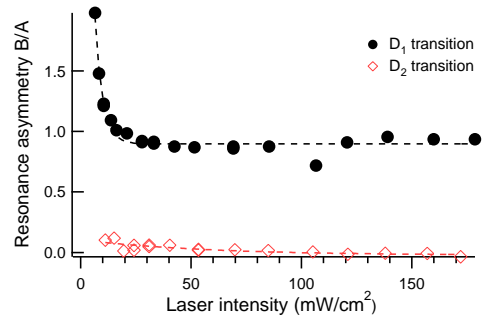}
\caption{
   \emph{N}-resonance asymmetry, $B/A$, as a function of laser
intensity, determined from fits of a combination of symmetric and
antisymmetric Lorentzian functions to measured \emph{N}-resonance
lineshapes; for $D_1$ (solid circles) and
   $D_2$ (open diamonds) transitions. Dashed lines are to guide the
   eye.}
\label{f.asymD1D2}
\end{figure}

Optically-probed atomic frequency standards commonly employ slow
modulation of the microwave drive and associated phase-sensitive
detection as part of the crystal oscillator lock-loop.  Hence an
asymmetric atomic resonance lineshape can induce systematic frequency
shifts proportional to the modulation parameters.\cite{CPTshifts} As
seen in Fig.~\ref{f.sampleD1D2}, ${}^{87}$Rb \emph{N}-resonances are
significantly more symmetric for the $D_2$ transition than for $D_1$,
which gives an important advantage for $D_2$ \emph{N}-resonance
operation. The relative asymmetry of the $D_1$ and $D_2$
\emph{N}-resonance lineshapes can be quantified by describing each
measured \emph{N}-resonance as a combination of symmetric and
antisymmetric Lorentzian functions:
\begin{equation}
T=T_{o}-\frac{A \, \Delta\nu / 2 \ + \ B \, \delta}{\Delta\nu^2/4 + \delta^2}
\label{e.asym}
\end{equation}
where $T$ is the measured probe field transmission as a function of
two-photon Raman detuning, $\delta$; and $A$, $B$, and $\Delta\nu$
are fit parameters that represent the amplitudes of the symmetric and
antisymmetric
Lorentzian components and the resonance linewidth. Fig.\
\ref{f.asymD1D2} shows the ratio of antisymmetric and symmetric
components, $B/A$, determined from our \emph{N}-resonance lineshape
measurements, as a function of laser intensity, indicating that the
$D_1$ \emph{N}-resonance is typically more than an order of magnitude
more asymmetric than the $D_2$ \emph{N}-resonance.

In conclusion, \emph{N}-resonances are
three-photon-absorption resonances that are a promising
alternative to CPT resonances for small atomic
frequency standards using alkali atoms. Here, we report an
experimental comparison of the $D_1$ and $D_2$ \emph{N}-resonances in
thermal ${}^{87}$Rb vapor.  We find that the $D_2$ \emph{N}-resonance
has better contrast but a broader linewidth than the $D_1$
\emph{N}-resonance, such that the \emph{N}-resonance quality factor
is comparable for the $D_1$ and $D_2$ transitions.  This result
implies a similar shot-noise-limit to  \emph{N}-resonance frequency
standard performance on the $D_1$ and $D_2$ transitions --- in stark
contrast with CPT resonances for which the quality factor is about an
order of magnitude worse for the $D_2$ transition than for $D_1$. In
addition, we find that the $D_2$ \emph{N}-resonance lineshape is
significantly more symmetric than the $D_1$ lineshape, indicating
that a $D_2$ \emph{N}-resonance frequency standard
will have reduced sensitivity to certain modulation-induced
systematic frequency shifts.  Thus, unlike for CPT resonances,
commercially available diode
lasers for the $D_2$ lines of Rb and Cs can likely be used without
compromising the performance of an \emph{N}-resonance frequency standard.

The authors are grateful to J.\ Vanier and M.\ Crescimanno for useful
discussions. This work was supported by ONR, DARPA, ITAMP and the
Smithsonian Institution. A. V. T. and V. I. Y. acknowledge support
from RFBR (grants no.\ 05-02-17086, 05-08-01389 and 04-02-16488). I.
Novikova's e-mail address: i.novikova@osa.org.
\\
$^*$ also at Lebedev Institute of Physics, Moscow, 
Russia.

\end{document}